\documentclass[twocolumn, prb,showpacs,preprintnumbers,amsmath,amssymb,superscriptaddress]{revtex4}

\usepackage{graphicx}
\usepackage{dcolumn}
\usepackage{bm}
\usepackage{upgreek}
\preprint{arXiv preprint}

\begin{document}

\title{Feedback-controlled laser fabrication of micromirror substrates}

\author{Benjamin Petrak}

\author{Kumarasiri Konthasinghe}%

\author{Sonia Perez}%

\author{Andreas Muller}%
 \email{mullera@usf.edu}
\affiliation{%
Dept. of Physics, University of South Florida, Tampa, FL 33620}

\date{\today}

\begin{abstract}
Short (40-200 $\upmu$s) single focused CO$_2$ laser pulses of energy $\gtrsim$100 $\upmu$J  were used to fabricate high quality concave micromirror templates on silica and fluoride glass. The ablated features have diameters of $\approx$20-100 $\upmu$m and average root-mean-square (RMS) surface microroughness near their center of less than 0.2 nm. Temporally monitoring the fabrication process revealed that it proceeds on a time scale shorter than the laser pulse duration. We implement a fast feedback control loop ($\approx$20 kHz bandwidth) based on the light emitted by the sample that ensures an RMS size dispersion of less than 5 \% in arrays on chips or in individually fabricated features on an optical fiber tip, a significant improvement over previous approaches using longer pulses and open loop operation.\end{abstract}

\maketitle

\section{Introduction}

During the past decade focused lasers have come to play an increasingly important role in the fabrication of photonic microstructures such as microlenses,\cite{beadie1995ssl, naessensa2003ffm} microresonators,\cite{collot1993vhw, armani2003uht, colombe2007saf} and waveguides. \cite{grobnic2006flf, psaila2006fli} Fabrication methods based on infrared lasers, notably CO$_2$ lasers, offer significant advantages over conventional etching-based methods, due to their ability to create ultralow loss dielectric interfaces by reflow of a glassy material. For example a microsphere resonator can be generated by melting an optical fiber with a CO$_2$ laser.\cite{collot1993vhw} Similarily, a microtoroid resonator can be fabricated by short laser exposure of an etched silica microdisk.\cite{armani2003uht} Microsphere and microtorroid resonators made this way sustain whispering gallery modes with record high ($>$ $10^8$) quality factor. Other breakthroughs using lasers in microfabrication include in-situ photoresist exposure for site-selective micropost resonator fabrication,\cite{dousse2008nps} as well as fabrication of Bragg gratings in fibers and on chips.\cite{davis1996wwg} These advances in microfabrication have significantly contributed to the rapid progress in quantum optics, quantum information science, cavity optomechanics and metrology witnessed in recent years.

A CO$_2$ laser can also directly ``machine'' a concave spherical feature on glass. As first demonstrated by Colombe et al., such a laser-ablated feature can serve as a substrate for a high-quality micromirror.\cite{colombe2007saf} Due to its smooth surface it is compatible with ultralow loss reflective dielectric coatings obtained by ion beam sputtering, and can thus be employed for constructing a high-finesse Fabry-Perot microcavity.\cite{hunger2010ffp, muller2010ufl} Such optical cavities are key components in quantum optics as they permit the generation of strong light-matter interactions within a small volume.\cite{vahala2003oma} The benefits derive from the ability of the CO$_2$ laser to both ablate and locally reflow the surface, giving it a microroughness typically on the order of 0.2 nm RMS or less. This figure is much lower than what is usually possible by dry or wet etching, although much progress is being made to improve methods based on etching.\cite{biedermann2010umm} Another advantage of the CO$_2$ laser ablation is the possibility of fabricating mirrors at the tip of optical fibers, a tedious task to accomplish with conventional lithography.\cite{steinmetz2006asf} However, although micromirrors fabricated using a CO$_2$ laser are now widely used in both atomic physics and condensed matter physics experiments, the current process suffers form a number of drawbacks making it impractical for large scale use or wherever reliability and reproducibility are crucial. The most severe drawback is the very large variation in size ($>$50\% RMS size dispersion) and precise location of fabricated templates, particularly small ones. Another is the redeposition of ablated material from near-neighbors in an array. These issues often preclude the fabrication of homogeneous arrays of high surface quality. Yet, reproducible arrays are essential for potential scaling of quantum information schemes that rely on microcavities on either atom chips,\cite{trupke2005mhf, purdy2008icq} or in semiconductor/dielectric hybrid systems.\cite{muller2009ceq}

Here we employ 40-200 $\upmu$s long focused CO$_2$ laser pulses and resolve the ablation process temporally; monitoring the light emitted by the sample as well as the transmitted light from a reference laser reveals the relevant time constants. The emitted light is used as an error signal in a feedback loop enabling the fabrication of arrays with an RMS size dispersion of less than 5\% for a range of parameters. We describe the fabrication method and report the characterization of templates made on various substrates. Detailed account is given of surface quality control and optical testing of fabricated templates.

\section{Experimental Setup}

\begin{figure}[h!]
\includegraphics[width=3.0in]{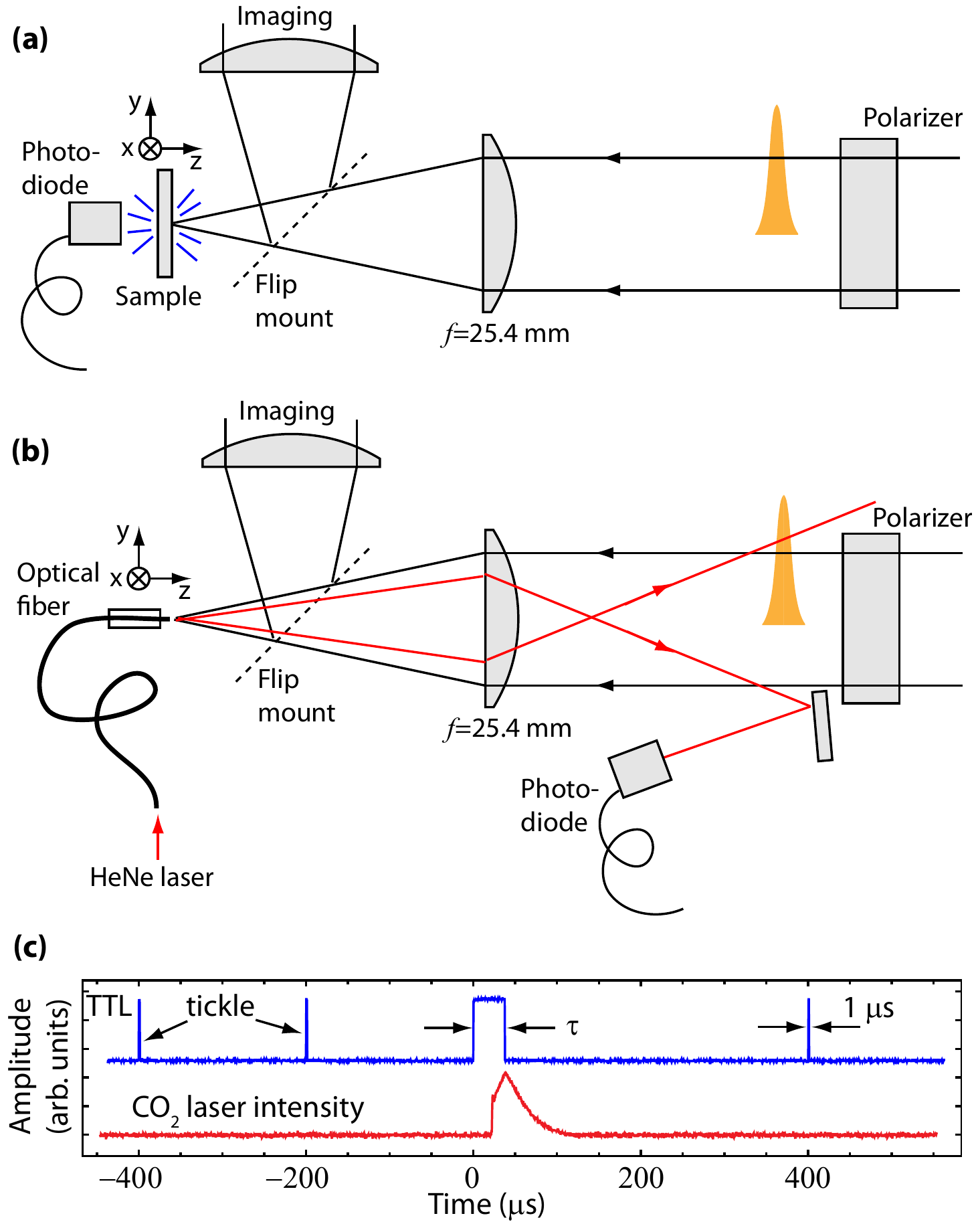}
\caption{\label{fig1} (Color online) (a) A CO$_2$ laser pulse is focused onto a planar fused silica substrate while the white light burst emitted is collected in transmission. (b) A HeNe laser is coupled into a single mode fiber and the CO$_2$ laser is focused on the fiber core. The back-propagating HeNe laser is picked off by a mirror and sent to a photodiode. At the instant the fiber tip is being ablated by the CO$_2$ laser, the HeNe laser divergence angle is changing which is detected as a change in intensity at the photodiode. (c) Electrical signal supplied to the laser (upper trace) and its actual intensity (lower trace).}
\end{figure}

The experimental setup is depicted in Fig. \ref{fig1}. All measurements described here employed a radio-frequency (RF) excited Synrad$\texttrademark$ laser as the source. The laser was linearly polarized with a wavelength near $\lambda$ = 10.6 $\upmu$m. The laser was spatially filtered by a 60 $\upmu$m pinhole at a focus, after going through a polarizer used to finely adjust its intensity. A ZnSe lens with focal length $f$ = 25.4 mm focused the laser tightly onto a sample. The sample was either a planar substrate or an optical fiber with its tip facing the incoming beam. The sample was positioned at the laser focus with a motorized x-y-z stage with the laser beam held fixed. A small flip mirror was temporarily inserted into the CO$_2$ laser beam path for imaging.  As a precise means of inserting and removing this imaging mirror (Fig. \ref{fig1} ) we attached it to the head of a recycled hard disk drive. Such a voice coil actuated head, previously used for making fast shutters,\cite{maguire2004hpl} electrically controlled the mirror position with minimal disturbance of the alignment and provided excellent repeatability. With the CO$_2$ laser spot size equal to $w=$4 mm at the lens, we estimate the spot size at the focus to be about $w=$ 20 $\upmu$m. 

We employ a field-programmable gate array (FPGA) to generate the correct TTL pulse sequence that controls the RF excitation of the laser plasma tube. ``Tickle" pulses 1 $\upmu$s long were applied at a frequency of 5 kHz to pre-ionize the plasma when not lasing, as specified by the manufacturer.\cite{synrad} Upon demand a single tickle pulse was replaced by a single square pulse of length $\tau$ that caused the generation of a laser pulse. The laser emission was characterized by an initial abrupt rise after a $\approx$ 20 $\upmu$s turn-on delay, a subsequent slow exponential rise, and finally an exponential decay once the electronic pulse was switched off [Fig. \ref{fig1} (c)]. The exponential rise and fall time constant of the pulse was on the order of 50 $\upmu$s. Despite this relatively slow response of the laser, the overall pulse duration could still be as small as about 40 $\upmu$s full width at half maximum (FWHM) because of the abrupt turn-on. For longer pulses we have observed that the tickle pulse immediately following a laser pulse actually caused unwanted laser emission; consequently we have removed it from the electronic triggering signal thereafter. When the pulse energy is at least about 100 $\upmu$J a concave spherical feature is being carved out by a single laser pulse. This process is accompanied by the emission of white light visible to the naked eye.

\section{Measurements}

\subsection{Temporally Resolving the Ablation Process}

In order to determine how the ablation process proceeds in time, we monitored either the burst of white light emitted by the sample [Fig. \ref{fig1} (a)], or the change in intensity that a back-propagating beam underwent during ablation [Fig. \ref{fig1} (b)]. For the former a detector was positioned behind a transparent sample while for the latter a HeNe laser was launched into a fiber on the tip of which the CO$_2$ laser was ablating a micromirror template; the intensity of the HeNe laser beam was monitored with a detector picking off the edge of the beam as it diverged after the ZnSe lens.

\begin{figure}[h!]
\includegraphics[width=2.8in]{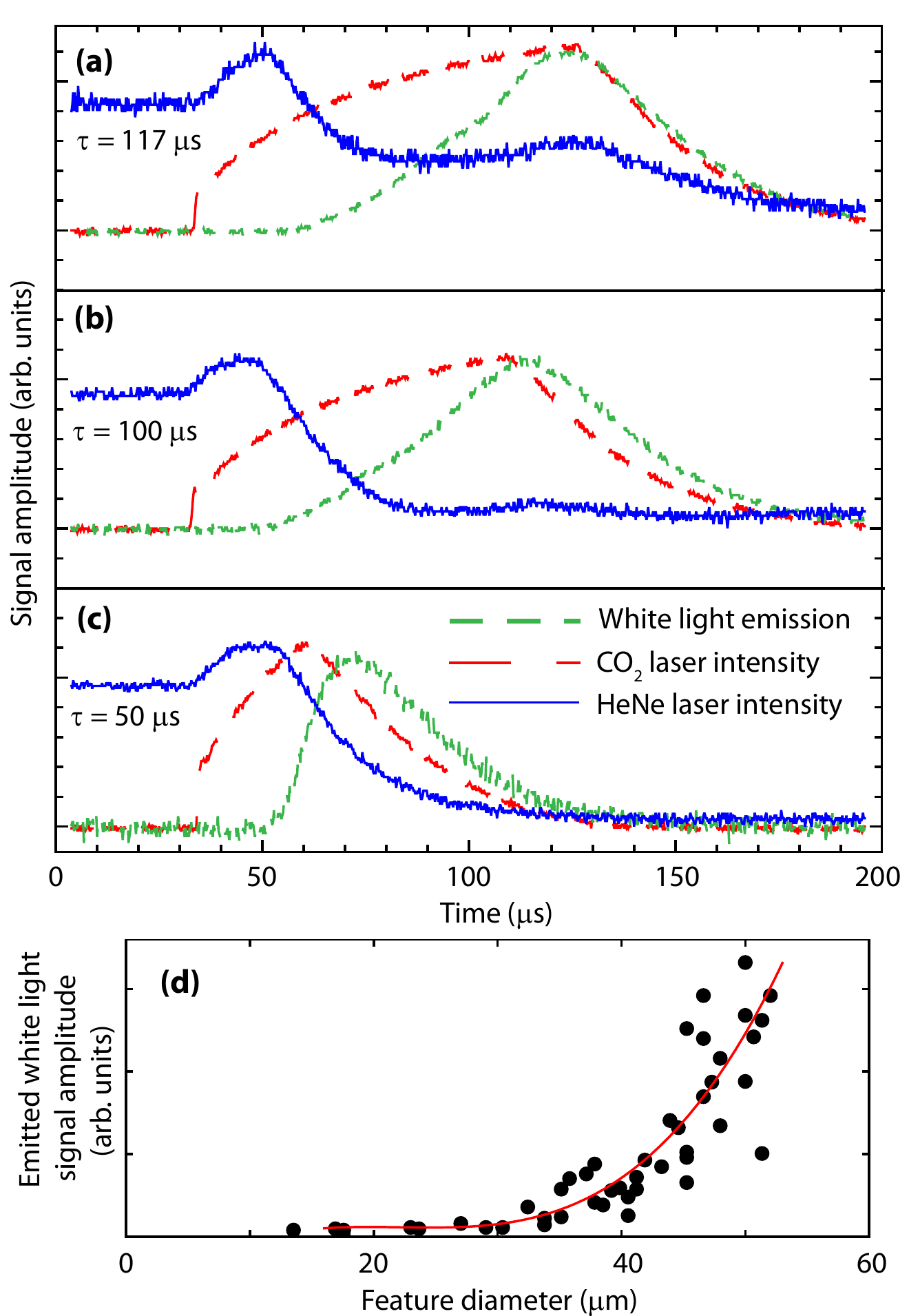}
\caption{\label{fig2} (Color online) (a-c) Normalized CO$_2$ laser intensity (long-dashed red trace), emitted white light intensity (short-dashed green trace), and back-propagating HeNe laser intensity (continuous blue trace) as a function of time for electronic pulse durations of $\tau$ = 117 $\upmu$s, 100 $\upmu$s, and 50 $\upmu$s respectively. The different signals have been measured separately and then plotted normalized. The experimental arrangement is that of Fig. 1(a,b). (d) Emitted white light intensity [peak amplitude of green short-dashed trace in Fig. 2(a-c)] plotted against fabricated feature diameter.}
\end{figure}

The result of these measurements is summarized in Fig. \ref{fig2} for various values of $\tau$. Within a few $\upmu$s of CO$_2$ laser switch on [long-dashed red line in Fig. \ref{fig2}(a-c) recorded with a pyroelectric photodiode], the material is beginning to be ablated, as seen by the sudden change in detected HeNe laser intensity [continuous blue line in Fig. \ref{fig2} (a-c)]. Some time later (typically tens of microseconds), light emission from the sample begins [short-dashed green line in Fig. \ref{fig2} (a-c)]. It persists even until after the laser intensity is starting to wane. When longer pulses are used, as in Fig. \ref{fig2} (a) and Fig. \ref{fig2} (b), a second phase of ablation occurs as the laser is switched off. As is shown in Fig. \ref{fig2} (d), the amount of white light emitted is monotonically increasing with the diameter of the fabricated feature.

\subsection{Closed Loop Control}

A large variability in the dimensions of fabricated features is often observed and is most extreme when using low energy pulses ($\approx$200 $\upmu$J). The variability can be so significant that for a given pulse energy one might obtain a 50 $\upmu$m, a 20 $\upmu$m diameter feature, or none at all. Although we do not know why these variations occur we have reasons to believe that they are exacerbated by the presence of dust or other contaminants on the surface. We have verified that variations of the energy supplied by the laser (typically on the order of $\pm$ 10\% partially due to fluctuations of laser turn on delay) can not account for the dramatic template size fluctuation. For many applications the smallest features are often the most desirable because they can be used to make Fabry-Perot microcavities with the smallest mirror spacing. Thus precise control over dimensions is important, even more so when making micromirrors at the tip of single mode fibers, the preparation of which is a time-consuming task. In order to address these difficulties we have devised and implemented a feedback control loop that allows us to switch off the laser once a desired amount of material has been ablated, as gauged by the amount of light emitted by the sample.

\begin{figure}[h!]
\includegraphics[width=3.0in]{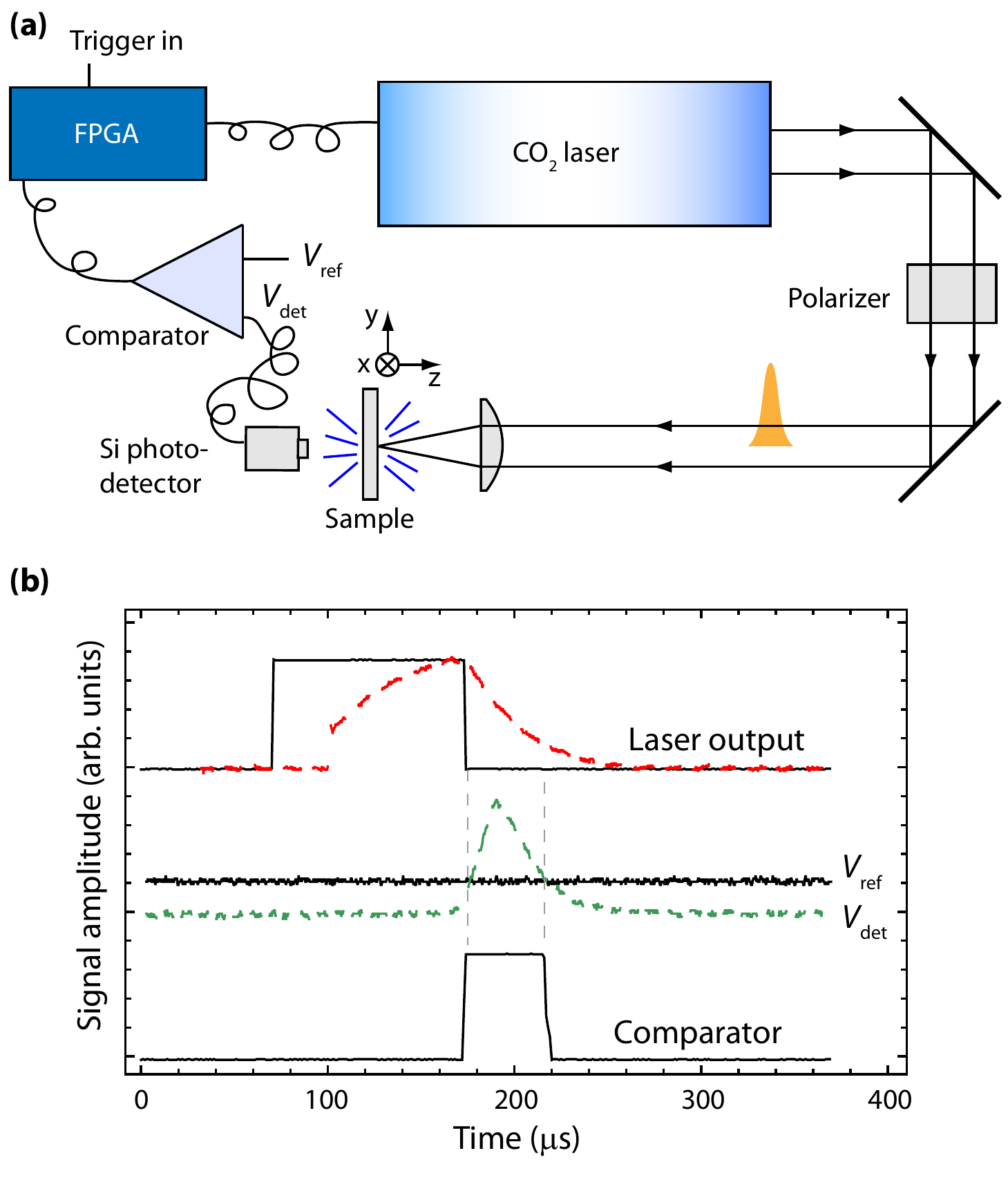}
\caption{\label{fig3} (Color online) (a) Schematic of feedback-controlled template fabrication. The CO$_2$ laser is switched on by the FPGA when a trigger signal is sent. The light emitted by the sample is detected by a Si photodetector giving rise to a voltage $V_{\rm det}$. A voltage comparator provides a TTL high signal when $V_{\rm det}>V_{\rm ref}$ where $V_{\rm ref}$ is an electronic reference voltage provided by a variable power supply. The comparator output is sent to the FPGA that in turn switches off the laser. (b) Temporal evolution of the feedback-controlled process. The electronic signal supplied to the laser (continuous top trace) and its actual output (dashed red trace) is turned high by the user and low when the comparator signal is high due to the emitted light exceeding the set threshold voltage, $V_{\rm ref}$.}
\end{figure}

Figure \ref{fig3}(a)  illustrates this feedback control implemented on the FPGA. After initial turn-on by a trigger, as soon as light is emitted by the ablated material it is detected by a photodetector giving rise to a voltage pulse $ V_{\rm det}$. This signal is compared to a reference voltage, $V_{\rm ref}$, using a fast voltage comparator that outputs a TTL pulse whenever $V_{\rm det}>V_{\rm ref}$. The comparator pulse is fed back to the FPGA which shuts down the laser emission. Thus, the overall effect is that when a certain amount of light has been received the ablation process is terminated, guaranteeing a minimal feature dimension but preventing the formation of an excessively large one. The laser shut off time is on the order of 40 $\upmu$s and light is emitted even after the electronic shutdown signal has arrived. Thus the bandwidth of the feedback loop is on the order of 20 kHz. Figure \ref{fig3} (b) illustrates how the process proceeds in time.

\subsection{Template Size Variation}

In order to estimate the average diameter of a fabricated feature, we automated the fabrication process and generated arrays with the same nominal exposure time (open loop operation) or the same nominal reference voltage (closed loop operation) for each feature within one array. The diameter of each feature was extracted from the image of the array recorded with an optical microscope. To automatically detect the center and diameter of each feature in an array we used third party software that is based on a Hough transform method.

\begin{figure}[h!]
\includegraphics[width=3.4in]{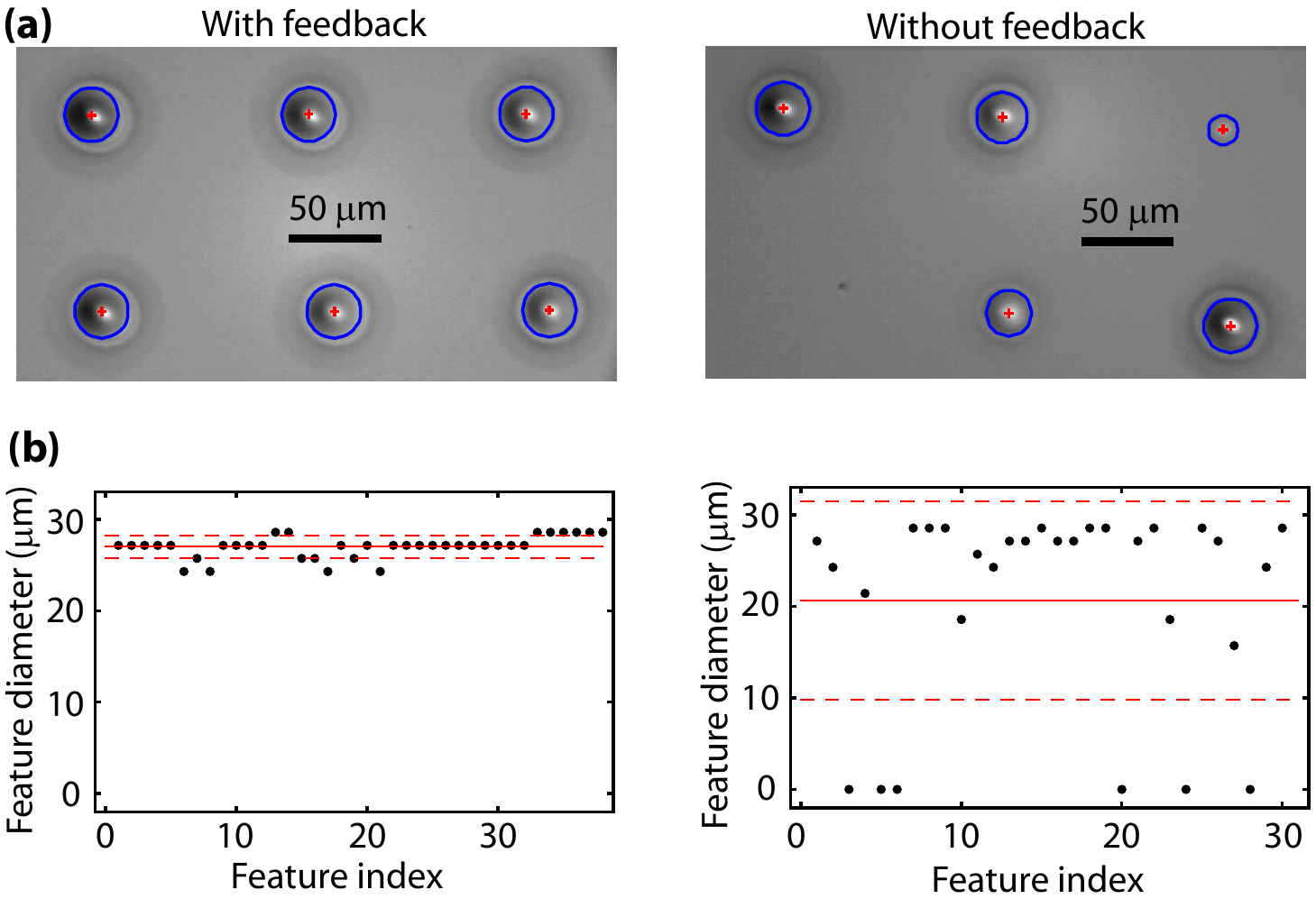}
\caption{\label{fig4} (Color online) (a) Optical microscope image of CO$_2$ laser fabricated micromirror substrates with (left) and without (right) feedback control. (b) Feature diameter for 38 (30) features with (left) and without (right) feedback control. When no feedback was used, the pulse energy was 270 $\upmu$J and its electronic trigger duration was 83.3 $\upmu$s. The misalignment between rows in the images is due to non-optimized motor settings. The blue circles were obtained from a Hough transform circle recognition program.}
\end{figure}

Figure \ref{fig4}  shows optical microscope images and extracted template diameters for fabrication with (left) and without (right) feedback control. As is obvious in the images, there is a significantly larger variability during open loop operation. We have quantified this variability by measuring the RMS feature diameter size fluctuation, defined as $\Delta D_{RMS} = \sqrt{\frac{1}{N}\sum_{i=1}^{N}{(D_i-\bar{D})^2}}$, where $\bar{D}$ is the mean diameter. We obtain $\Delta D_{RMS}/\bar{D} = $50\% for open loop operation (no feedback) and $D_{RMS}/\bar{D} = $4\% for closed loop operation (with feedback). This large difference, which is most pronounced for small features, highlights the advantage of the feedback controlled fabrication.

The feedback controlled microfabrication also works when the micromirror templates are fabricated at the tip of a silica fiber. In this case we collect the white light generated by the ablation process directly through the fiber (single mode or multimode). The feedback signal is obtained from a photodetector placed at the other fiber end that receives this white light. Although we have not performed a statistical analysis of the template size variation in this case we expect it to be identical to that obtained for planar fused silica substrates (Fig. \ref{fig4}).   

\subsection{Surface Microroughness}

\begin{figure}[h!]
\includegraphics[width=2.1in]{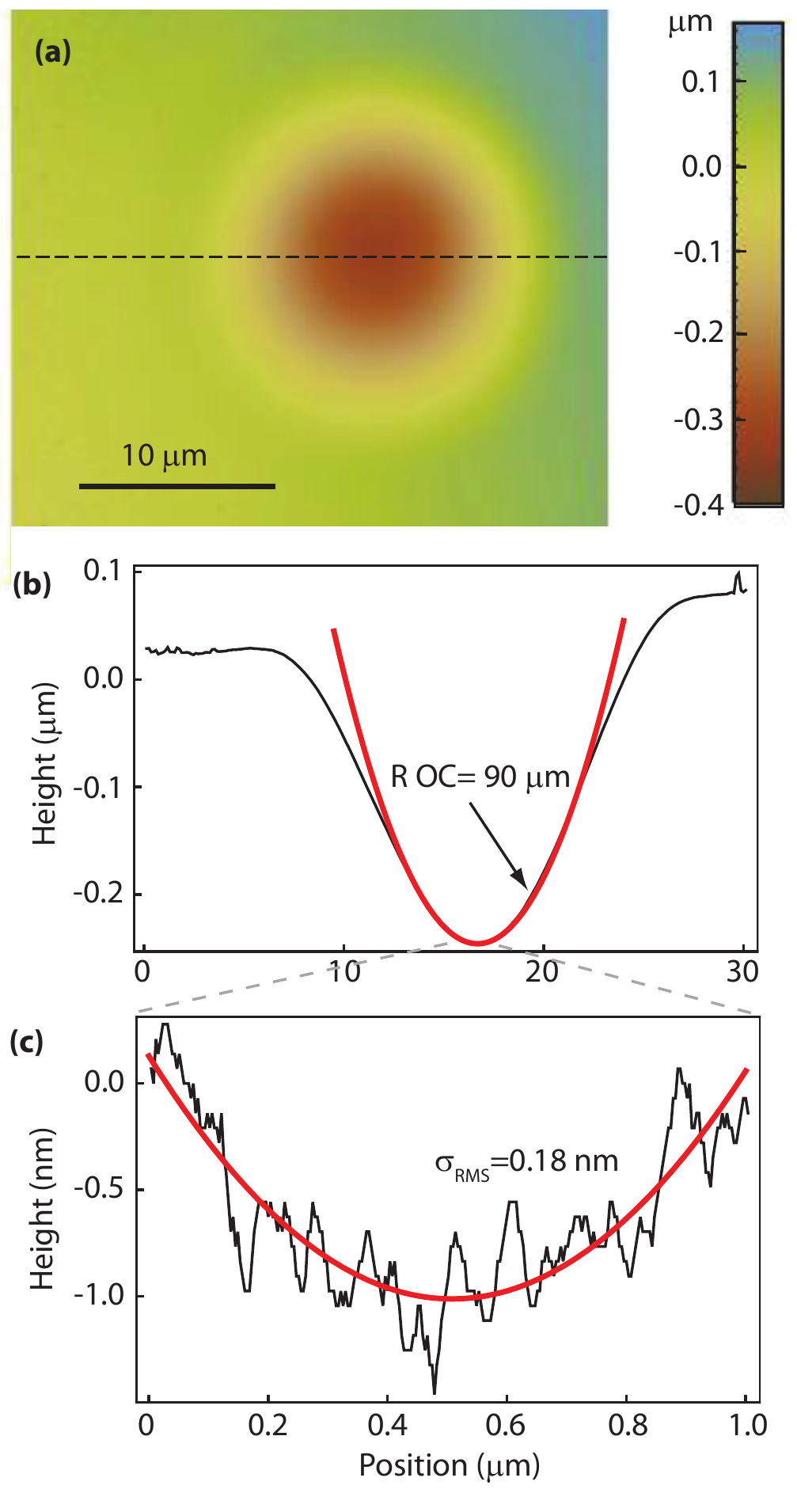}
\caption{\label{fig5} (Color online) (a) AFM image (unflattened) of a CO$_2$ laser fabricated micromirror substrate. (b) Cross-sectional profile extracted from image in (a). The depth, width, and radius of curvature (ROC) of the feature are respectively 0.3 $\upmu$m, 20 $\upmu$m, and 90 $\upmu$m. (c) Zoomed-in AFM scan over a 1 $\upmu$m distance near the center of the feature in (a).}
\end{figure}

An atomic force microscope (AFM) in tapping mode was employed to measure the cross-sectional profile of the micromirror substrates as well as their surface quality (Fig. \ref{fig5}). The height resolution of the instrument was determined to be $\sigma_{RMS}\approx$0.2 nm by topographic measurements of commercial superpolished substrates with nominal microroughness of 0.1 nm. Whenever specified, the surface microroughness was extracted as the RMS deviation from a parabolic fit to the height profile. Figure \ref{fig5}(a-c) shows the AFM image, cross-section and zoom-in cross-section of a fabricated feature on a planar fused silica sample. The cross section corresponds to the dashed line in the image. We have probed a total of 10 features, all of which yielded a value $\sigma_{RMS}\lesssim$0.2 nm uncorrected for the AFM resolution. Thus their actual surface microroughness may be smaller than 0.2 nm and if the coating would preserve this value then we would expect scattering losses of only about $(4\pi\sigma_{RMS}/\lambda)^2\approx 8(10)^{-6}$ at $\lambda$ = 900 nm.\cite{steinmetz2006asf} The finesse of a microcavity constructed with such mirrors would thus easily exceed $10^5$.

\subsection{Ferruled Fiber Fabrication}

\begin{figure}[h!]
\includegraphics[width=3.3in]{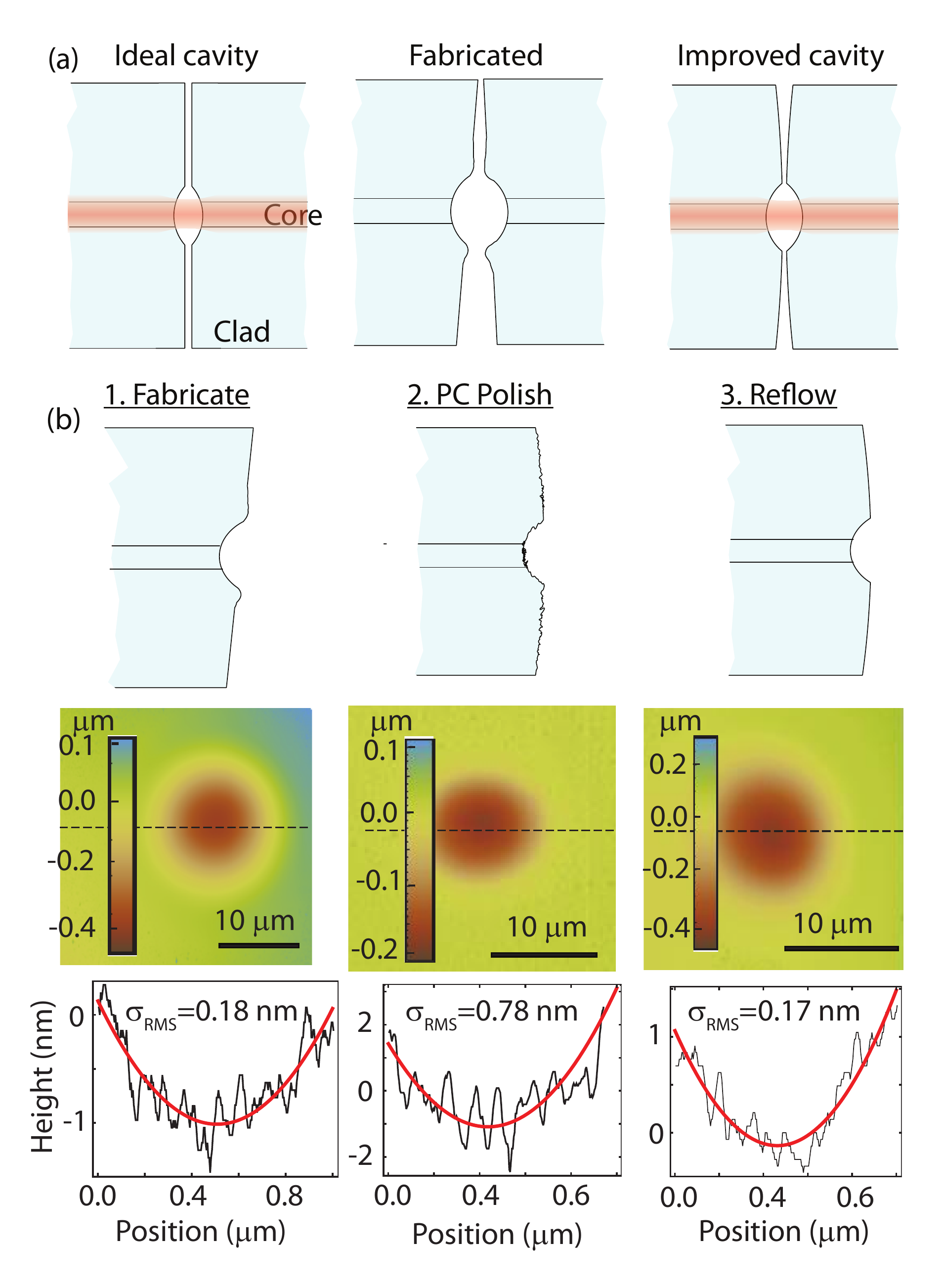}
\caption{\label{fig6} (Color online) Improved fabrication process on ferruled fibers. (a) The ideal cavity (left) differs from a real fabricated cavity (right) due to imperfections in fiber cleave and material redeposition during ablation. (b) A ferruled fiber that has been ablated is mechanically polished and re-exposed to a laser pulse of reduced energy to reflow the surface. The improved cavity enables a minimal mirror separation due to the PC polish. AFM images at each step of the procedure show the corresponding surface microroughness measurements.}
\end{figure}

Fabrication of micromirror templates on bare fibers suffer from several drawbacks such as residual cleave angle, poor heat dissipation, and other imperfections inherent to the laser fabrication process. These can be overcome when holding fibers in standard pre-radiused telecom ferrules and mechanically polishing them. This is useful when the goal is to make Fabry-Perot microcavities with very small spacing, i.e. very large free spectral range ($>$ 50 nm). The refined process we have used is illustrated in Fig. \ref{fig6}. Mechanical polishing is accomplished with 0.3 $\upmu$m lapping film until the desired amount of material has been removed. A second laser exposure is then applied to reflow the material after ultrasonically cleaning the substrate with methanol. A physical contact (PC) fiber polishing with a standard zirconia ferrule ensures minimal cavity length. We have performed AFM measurements at each step in the process and verified that the final surface quality is not degraded compared to the fabrication on unprocessed cleaved fibers or fused silica substrates. Ferruled fibers also make the final micromirrors more durable, less fragile than bare fibers, and compatible with other standard telecom connectors and holders. 

\subsection{Optical Testing}

\begin{figure}[h!]
\includegraphics[width=3.3in]{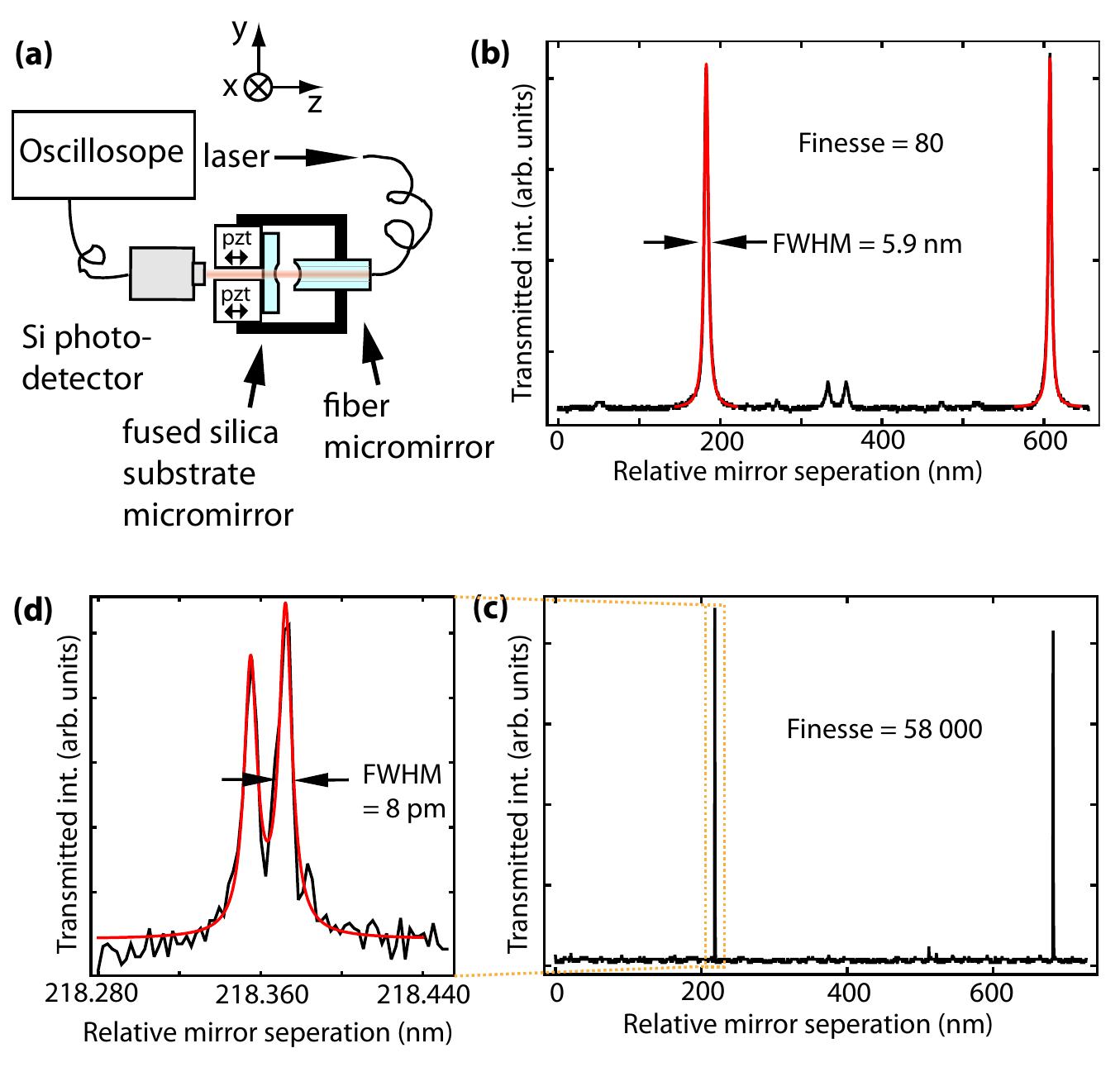}
\caption{\label{fig7} (Color online) Optical characterization of microfabricated and subsequently coated micromirrors. (a) A microcavity was set up by one fiber micromirror and one micromirror on a fused silica chip, separated by a small air gap ($\approx$ 10 $\upmu$m). Their mutual separation was continuously varied by a piezoelectric actuator. A cw laser was introduced into the fiber and the light transmitted through the cavity was collected by a photodetector. (b) Transmission of a cavity at $\lambda$=850 nm made of micromirrors coated with 20 nm of gold. (c) Transmission of a cavity at $\lambda$=940 nm made of micromirrors coated with a high-reflectivity dielectric coating. (d) Zoom into one of the resonances of (c). Note that the dashed zoom box is not drawn to scale.}
\end{figure}

We have tested the fabricated micromirror templates by applying a reflective optical coating, either a low-cost metallic coating consisting of 20 nm of gold, or a high-quality reflective dielectric coating by ion beam sputtering with a nominal reflectivity of $R$=99.995\% at 940 nm. The gold coating is useful for an initial verification of various parameters such as the template radius of curvature or for applications requiring only modest reflectivity, for example broadband filters. Micromirrors of the same dimensions were assembled into a cavity with a mirror separation controlled by a piezoelectric actuator [Fig. \ref{fig7}(a)]. One micromirror was on a plane substrate; the other was at the tip of a single mode fiber. While the separation between the two was varied, a continuous-wave (cw) laser was launched into the fiber and its transmission through the cavity was measured with a photodetector. The cavity finesse, $F=\pi/(1-R)$, was extracted as the ratio of the spatial separation between two contiguous longitudinal modes and their FWHM. For the gold coated mirror cavity the wavelength of the laser was $\lambda$=850 nm and the cavity finesse was $F$=80, corresponding to a reflectivity $R\approx$96\% [Fig. \ref{fig7}(b)]. In Fig. \ref{fig7}(c) the same measurement is shown but for the dielectrically coated micromirrors at $\lambda$=940 nm. There the finesse is larger than $F=$58 000, thus $R\approx$99.995\%. Fig. \ref{fig7}(d) shows a zoomed view of a fundamental transverse mode, which is actually split into two orthogonally-polarized resonances each with FWHM$\approx$8 pm. This splitting, commonly observed in high finesse microcavities is thought to originate in slight asymmetries of the micromirrors and/or in a birefringence of the coating \cite{muller2010ufl}. The finesse obtained compares well with that of state-of-the-art microcavities \cite{rempe1992mul, colombe2007saf} and we expect that it may be further increased without loss of transmission thanks to the extremely small surface microroughness of less than 0.2 nm.

\subsection{Substrate Material}

We have fabricated micromirror templates on various materials including float glass, BK7 glass, fused silica, ceramic Y$_3$Al$_5$O$_{12}$, and fluoride glass. Fabrication yielded good results only on fused silica and fluoride glass. For the other materials cracks typically formed and the surface quality was poor. Silica has excellent transmission properties up to $\lambda\approx$ 2 $\upmu$m. Fluoride glass, also available in the form of fibers from e.g. iRphotonics,\cite{IRphotonics} has good transmission properties in the mid infrared region up to $\lambda\approx$ 5 $\upmu$m.

\section{Conclusion}

In summary, short individual CO$_2$ laser pulses focused on a glass substrate create concave micromirror templates of high quality in a repeatable manner using the improved process described here. We are able to fabricate micromirrors with RMS variability inferior to 5\% and with a surface microroughness of less than 0.2 nm for most fabrication parameters. Further improvements of the size uniformity might be made by using a laser with a faster response so that the feedback control can be performed faster. Additional improvement might be obtained by integrating the intensity of the light received rather than by using a simple constant voltage threshold discrimination. Thus overall a highly precise fabrication method may be devised, possibly resulting in arrays of features with RMS size fluctuations inferior to 1\%.

Micromirrors that can be produced with the presently described optimized method may have uses for a wide variety of applications ranging from cavity quantumelectrodynamics in quantum optics experiments, tunable filters in telecommunication applications or in swept source optical coherence tomography, and microlasers. Laser micromirror fabrication can be applied to materials which cannot be processed easily with conventional methods but have otherwise desirable optical properties such as high infrared transmission. Precisely reproducible laser microfabrication may also be of great value for creating large arrays of concave microlenses at a very low cost.
$\newline$

We thank Phil Bergeron, Nick Djeu and Dennis Killinger for their contributions to this project. This work was supported in part by NSF REU program DMR-1004873.

\end{document}